\documentclass[english,preprint]{aastex}
\usepackage{mathptmx}
\usepackage[T1]{fontenc}
\usepackage[latin9]{inputenc}
\usepackage[letterpaper]{geometry}
\usepackage{textcomp}
\usepackage{amstext}
\usepackage{graphicx}

\makeatletter
\usepackage{times}
\makeatother

\usepackage{babel}

\begin{document}

\title{Disconnecting Open Solar Magnetic Flux}

\author{C.E. DeForest$^{*}$, T.A. Howard$^{*}$, and D.J. McComas$^{*\dagger}$}

\affil{$^{*}$Southwest Research Institute, 1050 Walnut Street Suite 300,
Boulder CO 80302}

\affil{$^{\dagger}$University of Texas at San Antonio, San Antonio TX 78249}

Disconnection of open magnetic flux by reconnection is required to
balance the injection of open flux by CMEs and other eruptive events.
Making use of recent advances in heliospheric background subtraction, we
have imaged many abrupt disconnection events. These events
produce dense plasma clouds whose distinctive shape can now be
traced from the corona across the inner solar system via heliospheric
imaging. The morphology of each initial event is characteristic of
magnetic reconnection across a current sheet, and the newly-disconnected
flux takes the form of a {}``U''-shaped loop that moves outward,
accreting coronal and solar wind material.

We analyzed one such event on 2008 December 18 as it formed and
accelerated at 20 m s$^{-2}$ to 320 km s$^{-1}$, expanding
self-similarly until it exited our field of view 1.2 AU from the
Sun. From acceleration and photometric mass estimates we derive the
coronal magnetic field strength to be $8\,\mu T$, 6$R_{\odot}$ above
the photosphere, and the entrained flux to be $1.6\times10^{11}Wb$
($1.6\times10^{19}Mx$).  We model the feature's propagation by
balancing inferred magnetic tension force against accretion drag.
This model is consistent with the feature's behavior and
accepted solar wind parameters.

By counting events over a 36 day window, we estimate a global event
rate of $1\, d^{-1}$ and a global solar minimum unsigned flux
disconnection rate of $6\times10^{13}Wb\, y^{-1}$ ($6\times10^{21}Mx\,
y^{-1}$) by this mechanism. That rate corresponds to $\sim-0.2\, nT\,
y^{-1}$ change in the radial heliospheric field at 1 AU,
indicating that the mechanism is important to the heliospheric
flux balance.

\section{Introduction}

The solar dynamo generates magnetic flux inside the Sun, whch is transported
outward and emerges through the Sun\textquoteright{}s surface into
the corona. Magnetic loops build up \textquotedblleft{}closed\textquotedblright{}
magnetic flux (connected to the Sun at both ends) in the corona. Some
of these closed loops subsequently \textquotedblleft{}open\textquotedblright{}
into interplanetary space \textendash{} that is, they are connected
to the Sun at only one end with the other extending to great distances
in the heliosphere or beyond. Owing to the very high electrical conductivity,
open magnetic flux is frozen into the solar wind and carried out with
it. The magnetized solar wind expands continuously outward from the
Sun in all directions, filling and inflating our heliosphere and protecting
the inner solar system from the vast majority of galactic cosmic rays.
The balance between the opening and closing of magnetic flux from
the Sun is thus critical and fundamental both to the solar wind and
to the radiation environment of our solar system.

Magnetic flux opens when coronal mass ejections (CMEs) erupt through
the corona, carrying previously closed magnetic loops beyond the
critical point where the solar wind exceeds the Alfv\'en speed
(typically <$20\, R_{\odot}$) and can no longer return to the
Sun. CMEs were first studied in OSO-7 and Skylab observations of the
corona (e.g.\citealt{Tousey1973,Gosling1974,Hundhausen1993}), and
since then continued work has provided an increasingly detailed
picture of these transient magnetic structures both during their
formation and ejection, and as they continue to evolve and interact
with the solar wind.  

Long lasting, radial \textquotedbl{}legs\textquotedbl{} are often
observed along the flanks of a CME and persisting behind it. These
legs are generally interpreted as evidence for at least some continued
magnetic connection of CMEs back to the Sun and hence the opening
of new magnetic flux with CME ejections. That picture is further supported
by observatation, \emph{in situ}, of beamed suprathermal halo electrons
streaming in both directions along the local interplanetary magnetic
field (IMF) during the passage of an interplanetary CME (ICME) cloud
(e.g. \citealt{Gosling1990,Gosling1993} \& references therein), which
are commonly interpreted as signatures of direct connection of the
ICME magnetic field to the solar corona in both directions, and hence
of newly opening magnetic flux. However, less is known about the equally
necessary process of disconnection that must be present to remove
newly opened flux and prevent the IMF from growing without limit.

Because of the continual opening of magnetic flux through CMEs, McComas
and coworkers in the early 1990s pursued a series of studies to determine
how magnetic flux could be closed back off and avoid a so-called magnetic
flux \textquotedblleft{}catastrophe\textquotedblright{} of ever increasing
magnetic field strength in the interplanetary magnetic field (\citealt{McComas1995}\&
references therein). The amount of open magnetic flux in interplanetary
space can be approximated with the \textquotedbl{}total flux integral\textquotedbl{}
which removes the effects of variations in the solar wind speed in
determining the amount of magnetic flux crossing 1 AU (\citealt{McComas1992a}).
Using this integral, McComas et al. (\citeyear{McComas1992a,McComas1995})
showed that if all counterstreaming electron events represent simply
connected opening magnetic loops, then for solar maximum CME rates,
the amount of flux crossing 1 AU would double over only \textasciitilde{}9
months. For flux rope CMEs, significantly more magnetic flux may be
observed in the loops crossing 1 AU than what remains attached to
the Sun along the CMEs\textquoteright{} legs; however, it must be
stressed that if CMEs retain any solar attachment whatsoever, the
flux catastrophe will ultimately occur in the absence of some other
process to close off previously open fields.

Of course a magnetic flux catastrophe is not observed in the solar
wind and, in fact, the overall magnitude of the IMF and amount of
open flux seems to vary over the solar cycle. For cycle 21, the average
magnitude varied by \textasciitilde{}50\% (\citealt{Slavin1986})
while the total flux integral varied by \textasciitilde{}60\% (McComas
et al. \citeyear{McComas1992a,McComas1992b}), with maxima shortly
after solar maximum and minima shortly after solar minimum \citealt{McComas1994}.
Since these studies, the solar wind has gone through a prolonged (multi-cycle)
reduction in both solar wind power (the dynamic pressure of the solar
wind that ultimately inflates the heliosphere) (\citealt{McComas2008})
and magnetic field magnitude (\citealt{Smith2008}). The lack of a
flux catastrophe, solar cycle variation and now long-term reduction
in the open magnetic flux from the Sun all show that there must be
some process for closing off previously open field regions and returning
magnetic flux to the Sun.

Magnetic reconnection plays an important role in regulating the topology
of solar magnetic flux, however, once the top of a loop passes the
critical point, its magnetic flux remains open until some other process
occurs to close it off below the critical point. That is, reconnection
above the critical point can only rearrange the topology of open magnetic
flux in the heliosphere \textendash{} only reconnection between two
oppositely directed (inward and outward field) regions of open flux
close to the Sun can close off previously open magnetic flux. The
most obvious method of reducing the amount of magnetic flux open to
interplanetary space is via reconnection between oppositely directed,
previously open field lines (McComas et al. \citeyear{McComas1989}),
which creates closed field loops that can return to the Sun and the
release of disconnected U-shaped field structures into interplanetary
space. An example of such a coronal disconnection event was shown
by \citealt{McComas1991} using SMM coronagraph images from 1 June
1989. An even older example of a likely coronal disconnection event
can be found as far back as the 16 April 1893 solar eclipse (e.g.,
\citealt{Cliver1989}), where sketches (data in 1893) made in time
ordered sequence from Chile, Brazil, and Senegal, indicate the outward
motion of a large U-shaped structure (\citealt{McComas1994}).

For the opening and closing of the solar magnetic flux to maintain
some sort of equilibrium, there must be some type of feedback between
these two processes. McComas et al. (\citeyear{McComas1989,McComas1991})
suggested that this feedback occurs through transverse magnetic pressure
in the corona, where the expansion of newly opened field regions must
enhance transverse pressure and compress already open flux elsewhere
around the Sun. When enough pressure builds up, reconnection between
oppositely direct open flux would reduce the pressure and amount of
open flux. The sequence of images from the 27 June 1988 coronal disconnection
event, in fact showed just such a compression, indicated by the deflection
of the streamers in the corona, just prior to and appearing to precipitate
the coronal disconnection event. Another line of supporting evidence
was provided by numerical simulations (\citealt{Linker1992}), which
indicated that increased magnetic pressure could lead to reconnection
across a helmet streamer and the release of disconnected flux. \citet{Schwadron2010}
recently reexamined the flux balance issue in light of the anomalously
long solar minimum between cycles 23 and 24 and modeled the level
of magnetic flux in the inner heliosphere as a balance of that flux
injected by CMEs, lost through disconnection, and closed flux lost
through interchange reconnection near the Sun.

\citet{McComas1992c}conducted a statistical study of three months
of SMM coronagraph observations (\citet{Hundhausen1993}) to assess
the frequency of coronal disconnection events. These authors found
that while the initial survey (\citealt{StCyr1990}) found no obvious
disconnections, six of the 53 transient events during this interval
(11\%) showed some evidence of disconnection in more than one frame
and 13 (23\%) showed a single frame with an outward \textquotedbl{}U\textquotedbl{}
or \textquotedbl{}V\textquotedbl{} structure. Given the imaging and
analysis technology of the day, McComas et al. (1992c) concluded that
magnetic disconnection events on previously open field lines may be
far more common than previously appreciated. With today\textquoteright{}s
imaging and exceptional analysis capabilities, the question of coronal
disconnection events should finally be resolvable. 

For this study we used image sequences, collected by the SECCHI
(\citealt{HowardRA2008}) instrument suite on board NASA's
\emph{STEREO-A }spacecraft, of Thomson-scattered sunlight from free
electrons in the interplanetary plasma. The observations span from the
deep solar corona to beyond 1 AU at elongation angles of up to
$70^{\circ}$ from the solar disk; this continuous observation is
enabled by recently developed background subtraction techniques
(\citealt{DeForest2011}) operating on the \emph{STEREO} data. The
signature U-shaped loops of disconnected plasma are far clearer in the
processed heliospheric data far from the Sun than in the coronagraph
data close to the Sun, and we detect 12 characteristic departing
{}``V'' or {}``U'' events in 36 days -- far more than the expected
number based on scaling the results of McComas et al. (1992c). For
this initial report, we focus on quantitative analysis of a single
event. In Sections 2.1-2.7, we describe the observations and calculate
the geometry, the mass evolution, and (by assuming the U-loop is
accelerated by the tension force) the coronal magnetic field sand
entrained flux in the disconnecting structure. As a plausibility
check, we explore the tension force scenario and its consequences for
the long-term evolution of the feature, and find that the scenario is
consistent with accepted values for the solar wind density and
speed. In Section 3, we discuss broader consequences of the
observation, including estimating the disconnection rate based on the
number of similar events in our data set, and discuss implications for
the global magnetic flux balance.

\section{Observations}

The SECCHI suite on STEREO was intended to be used as a single integrated
imaging instrument (e.g. \citealt{HowardRA2008}). It consists of
an EUV imager (EUVI) observing the disk of the Sun, and four visible
light imagers (COR-1, COR-2, HI-1, and HI-2) with progressively wider
overlapping fields of view, to cover the entire range of angles between
the solar disk and the Earth. The visible light imagers view sunlight
that has been Thomson scattered off of free electrons in the corona
and interplanetary space; the theory of Thomson scattering observations
has been recently reviewed by \citet{HowardTappin2009a}. We set out
to view coronal and heliospheric events in the weeks around 2008 December,
using newly developed background subtraction techniques to observe
solar wind features in the HI-1 and HI-2 fields of view (\citet{DeForest2011,HowardDeForest2011}).
In the initial 36 day data set we prepared, we observed 12 disconnection
events identified by a clear {}``V'' or {}``U'' shaped bright
structure propagating outward in the heliosphere. We chose a particuarly
clearly presented one, which was easily traceable to its origin in
the low corona on 2008-Dec-12 at 04:00, for further detailed study.

\subsection{Image Preparation}

Data preparation followed standard and published techniques. For COR-1
and COR-2, we downloaded Level 1 (photometrically calibrated) data,
and further processed them by fixed background subtraction: we acquired
images for an 11-day period, and found the 10 percentile value of
each pixel across the entire 11-day dataset. This image was median
filtered over a 5x5 pixel window to generate a background image that
included the F corona, any instrumental stray light, and the smooth,
steady portion of the K corona. Subtracting this background from each
image yielded familiar coronal images of excess feature brightness
compared to the smooth, steady background. We performed one additional
step: motion filtration to suppress stationary image components. This
step matches the motion filtration step used for HI-1 and HI-2 (below),
and suppresses the stationary streamer belt while not greatly affecting
the moving features under study.

The heliospheric imagers required further processing to remove the
starfield, which is quite bright compared to the faint Thomson scattering
signal far from the Sun in the image plane. We processed the STEREO-A
HI-2 data as described by \citet{DeForest2011}. The HI-1 data used
a similar process adapted to the higher background gradients in that
field of view and described by \citet{HowardDeForest2011}.

All the imagers yielded calibrated brightness data in physical units
of the mean solar surface brightness ($ $$ $ $B_{\odot}=2.3\times10^{7}W\, m^{-2}\, SR^{-1}$).
Because of the wide field of view, as a subsequent processing step
we distorted the images into azimuthal coordinates, in which one coordinate
is azimuth (solar position angle) in the image plane and the other
is either elongation angle $\epsilon$ ({}``radius'' on the celestial
sphere) or its logarithm. The latter projection, if scaled properly,
is conformal: it preserves the shape of features that are small compared
to their distance from the Sun. To equalize brightness, we applied
radial filters to the images for presentation, with either a $\epsilon^{3.5}$
scaling (for coronal images) or a $\epsilon^{3}$ scaling (for heliospheric
images).

\begin{figure}[tbh]
\begin{centering}
\includegraphics[width=6in]{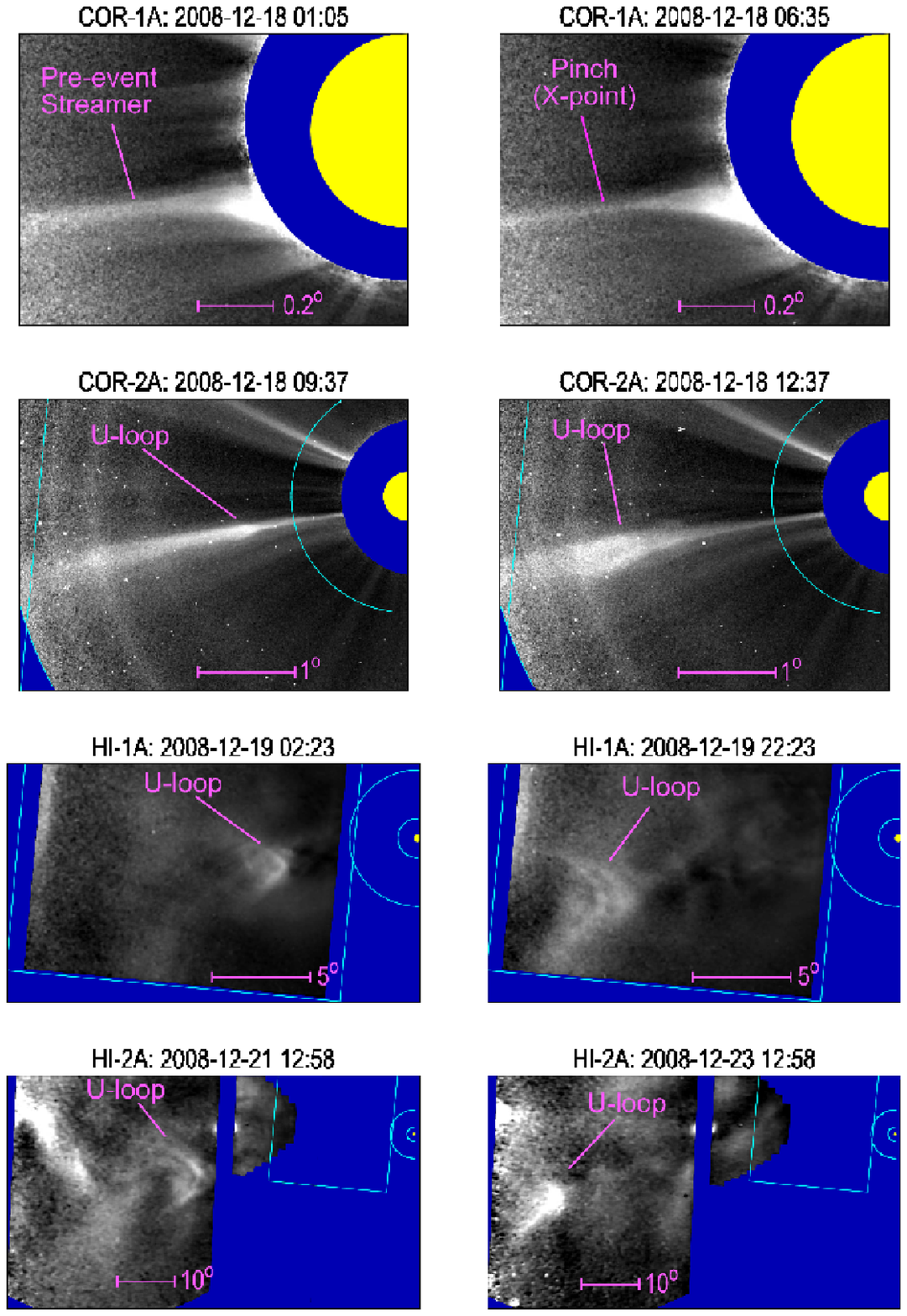}
\par\end{centering}

\caption{\label{fig:4-panels}The disconnection event of 2008 Dec 18 in context
- 8 still images showing formation and evolution of the U-loop: left
to right, top to bottom.}
\end{figure}

Figure \ref{fig:4-panels} shows snapshots of the disconnected plasma
and associated cusp, as observed by four separate instruments over
the course of four days as it propagated outward. Shortly after 2008
Dec 18 04:00, the streamer belt at $160^{\circ}$ ecliptic azimuth
($20^{\circ}$ CCW of the Sun-Earth line) pinched and separated, forming
a {}``U'' loop that retracted outward, with a trailing cusp, over
the course of the following three days. The feature remained visible
in Thomson scattered light because of plasma scooped up during the
early acceleration period in the lower corona: this plasma remained
denser than the surrounding medium, yielding a bright feature throughout
the data set. The disconnected plasma completely missed the ecliptic
plane and was therefore not observed \emph{in situ }by any of the
near-Earth or STEREO probes.

\subsection{Observing Geometry and 3-D structure }

\begin{figure}[tbh]
\begin{centering}
\includegraphics[width=5in]{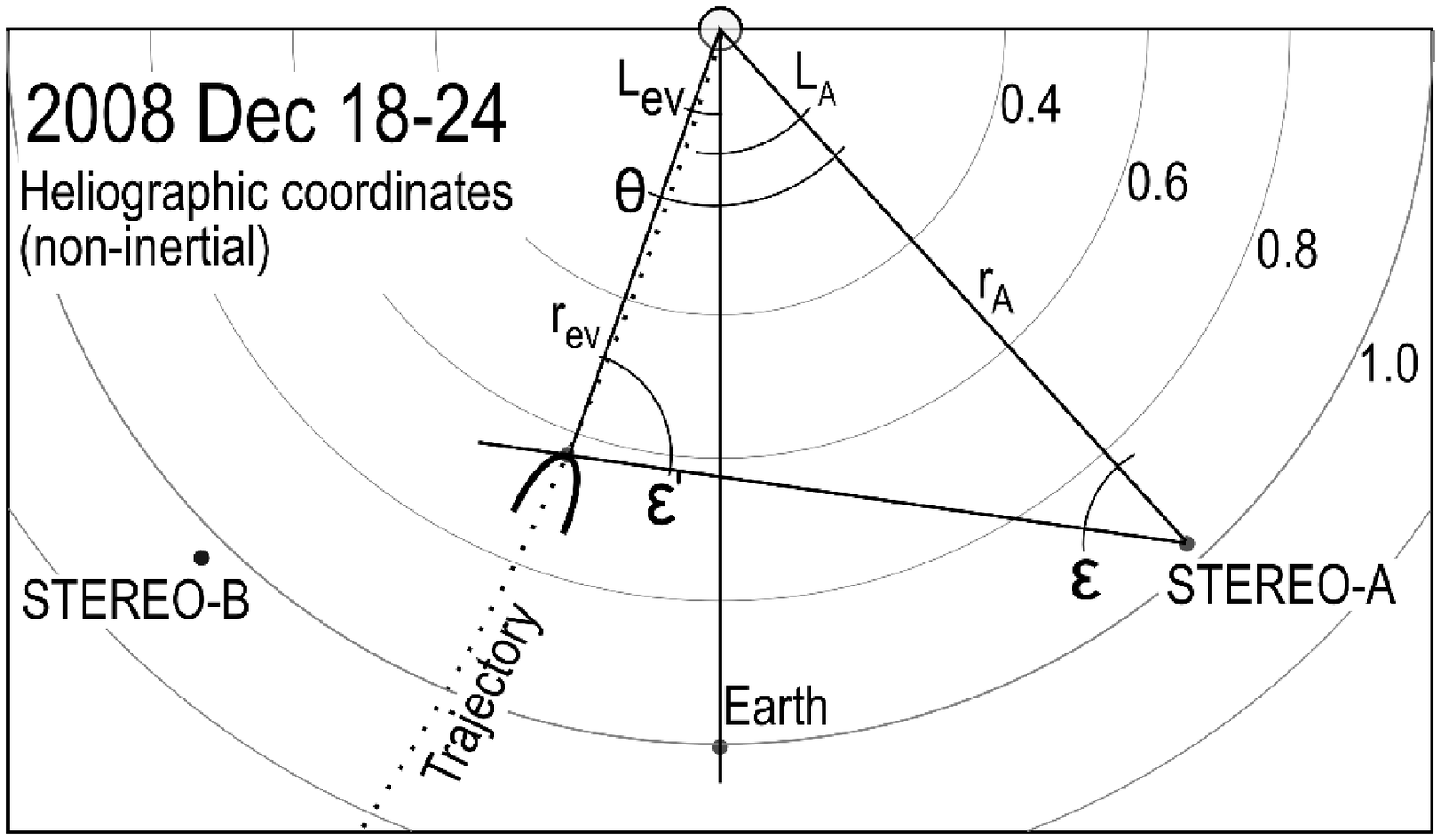}
\par\end{centering}

\caption{\label{fig:Observing-geometry}Observing geometry in the ecliptic
plane on 2008 Dec 18 - 2008 Dec 22}
\end{figure}

The observing geometry for the 2008 Dec 18 event is shown in Figure
\ref{fig:Observing-geometry}, from an overhead (northward out-of-ecliptic)
point of view co-rotating with the STEREO-A orbit. The event departure
angle was measured both using direct triangulation between the coronagraphs
in STEREO-A and STEREO-B. We used the triangulation method described
by \citet{HowardTappin2008}. Although the disconnection event is
small compared to most CMEs, subtending just a few degrees in latitude,
it is still large enough to cast doubt on the simple triangulation
results, so we also used {}``TH model'' semi empirical transient
event reconstruction tool (\citealt{TappinHoward2009}) to extract
the departure angle. TH was developed to reconstruct CME leading edge
({}``sheath'') overall envelope and propagation speed, but is also
applicable to smaller transient events such as this one. Details,
applications, and limitations of the TH model are further described
by \citealt{TappinHoward2009} and by Howard \& Tappin (\citeyear{HowardTappin2009b,HowardTappin2010}).
Departure longitude was measured to be $-10^{\circ}\text{\textpm}5^{\circ}$
in heliographic coordinates, with an estimated event width of under
$5^{\circ}.$

We took the disconnected feature's trajectory to have constant radial
motion (the {}``Fixed-$\Phi$ aproximation'') in the solar inertial
frame -- this leads to the slightly curved aspect to the trajectory
in the co-rotating heliographic ecliptic frame, which maintains the
prime meridian at the Earth-Sun line. Figure \ref{fig:Observing-geometry}
shows an out-of-ecliptic projected view of the observing geometry,
including construction angles and distances used in Section \ref{sub:Acceleration-profile}
for trajectory calculations.

\subsection{Feature Evolution}

\begin{figure}[tbh]
\begin{centering}
\includegraphics[width=1\textwidth]{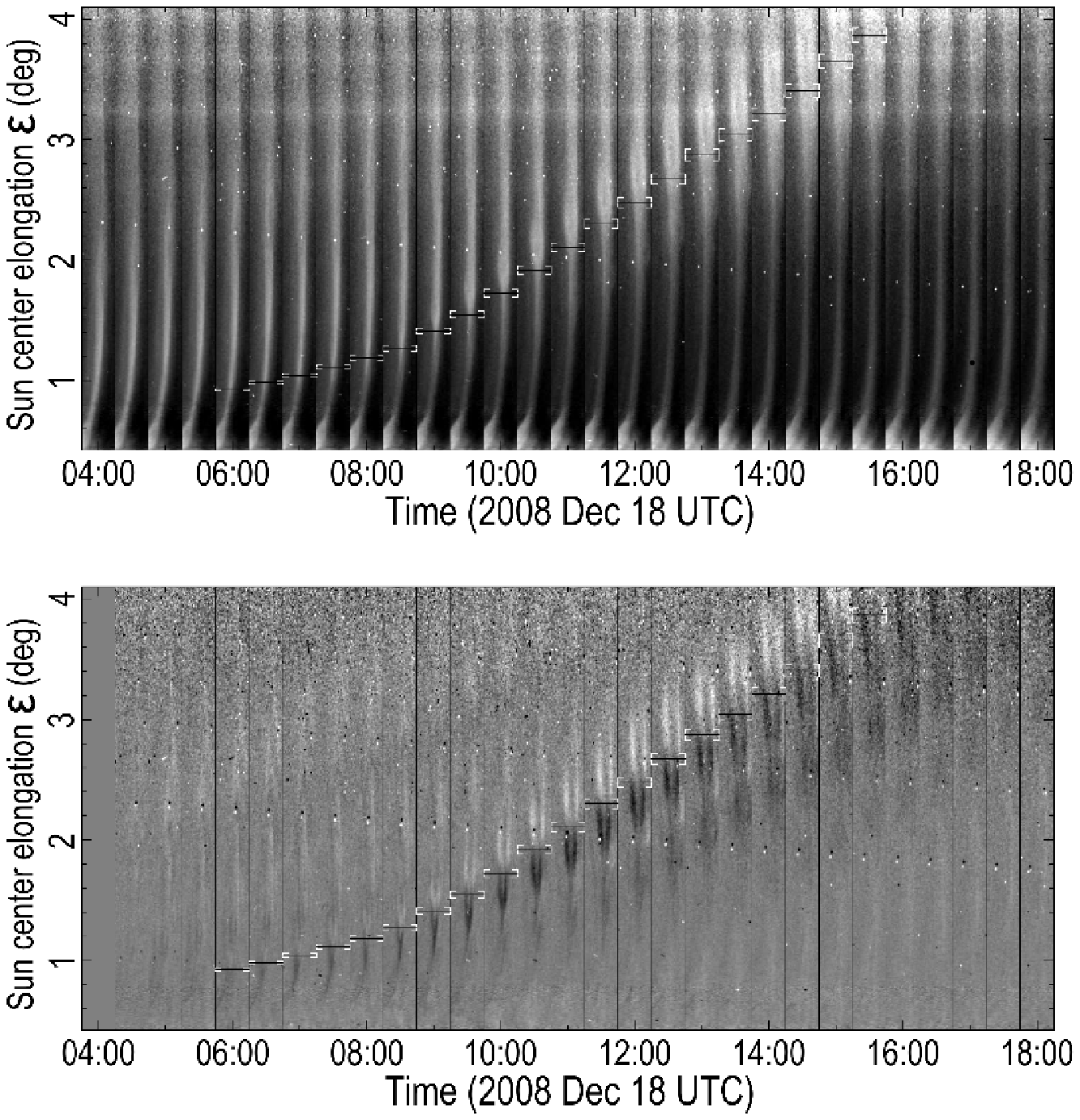}
\par\end{centering}

\caption{\label{fig:liftoff}Formation and early acceleration of the 2008 Dec
18 disconnection event through the STEREO-A COR-1 and COR-2 fields
of view The trailing edge of the event is marked, with error bars
based on feature identification. TOP: direct excess-brightness images
show feature formation and overall structure. BOTTOM: running-difference
images show detail. These stack plots include a small image of the
feature at each sampled time to show evolution. Intensities are scaled
with $\epsilon^{3.5}$ to equalize brightness vs. height. The individual
images have been resampled into linear azimuthal (radial) coordinates,
and the horizontal range is $160^{\circ}-174^{\circ}$ of azimuth.
Note that this projection does not preserve aspect ratio: despite
appearances, the event widens as it rises.}
\end{figure}

\begin{figure}[tbh]

\begin{centering}
\includegraphics[width=1\textwidth]{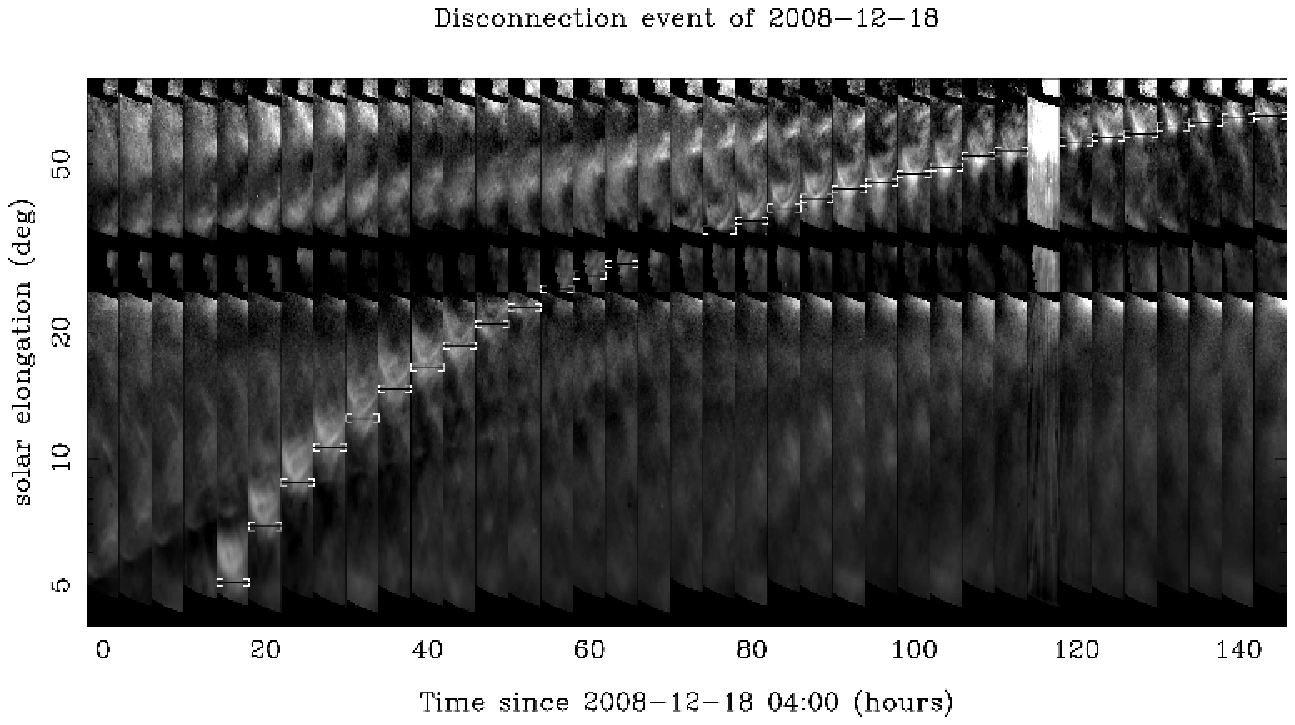}
\par\end{centering}

\caption{\label{fig:Propagation}Propagation and evolution of the 2008 Dec
18 disconnection event through the STEREO-A HI-1 and HI-2 fields of
view. The trailing edge of the event is marked, with error bars based
on feature identification. These stack plots include a small image
of the feature at each sampled time to show evolution. The individual
images have been resampled into logarithmic azimuthal (radial) coordinates,
and the horizontal range is $162^{\circ}-178\text{\textdegree}$of
azimuth. This projection is conformal, so the shape of the feature
is preserved in each image. Intensities are scaled with $\epsilon^{3}$
to equalize brightness vs. height. Note self-similar expansion: the
angular width and shape of the feature are preserved.}

\end{figure}

To analyze the feature's evolution across a two-order-of-magnitude
shift in scale over its observed lifetime, we transformed the processed
\emph{STEREO-A} source images into local heliographic radial coordinates
-- i.e. zero azimuth is due solar West from the viewpoint of \emph{STEREO-A},
with azimuthal coordinate increasing clockwise around the image plane;
this follows early work by \citet{DeForestPlunkettAndrews2001} in
imaging polar plumes. Distances from Sun center are recorded as elongation
angle $\epsilon$ from the center of the Sun, as a reminder of the
angular nature of the wide-field observations. To avoid aliasing in
the resampling process, we resampled the images using the optimized
resampling package described by \citet{DeForest2004}. Figures \ref{fig:liftoff}
and \ref{fig:Propagation} show the liftoff and propagation of the
feature across 65 degrees of elongation from its origin in the solar
streamer belt. Both figures have a radial gain filter applied to equalize
the feature's brightness, which varies by over seven orders of magnitude:
from $1.5\times10^{-9}B_{\odot}$ in the low streamer belt at 2008
Dec 18 04:30 to $5.7\times10^{-17}B_{\odot}$ six days later, at $\epsilon=65^{\circ}$. 

The bright feature takes the classic wishbone shape of reconnecting
field lines emerging from a current sheet (e.g. Chapter 4 of \citealt{PriestForbes2000}).
The aspect ratio of the wishbone may be estimated by dividing the
vertical height from cusp to the top of the visible horns, by the
width between the horns. This aspect ratio varies from \textasciitilde{}10:1
when the horns are first clearly resolved near 2008 Dec 18 08:00,
to approximately 2:1 some four hours later and 1:1 by 2008 Dec 19
04:00 -- one full day after the first pinch is observed in the streamer
belt. After 2008 Dec 19, the feature expands approximately self-similarly
as it propagates, subtending approximately $16^{\circ}$ of azimuth
and not changing its aspect ratio throughout the rest of its trajectory. 

Note that aspect ratio is \emph{not} preserved by the linear azimuthal
mapping used in Figure \ref{fig:liftoff}, which was selected to show
the early acceleration clearly; aspect ratio is preserved by the (conformal)
logarithmic mapping used in Figure \ref{fig:Propagation}, which shows
nearly self-similar expansion in the image plane despite perspective
effects that come into play above about $\epsilon=30^{\circ}$

The scaling of brightness is reassuring because, in a uniformly propagating
wind with no acceleration, density must decrease as $r^{-2}$ and
feature column density must thus decrease as $r^{-1}$, while illumination
decreases as $r^{-2}$, so feature brightness is expected to decrease
as $r^{-3}$. The fact that brightness levels do not change much across
$ $Figure \ref{fig:Propagation}, which is scaled by $\epsilon^{3}$,
suggests that the disconnected flux and material entrained in it are
indeed propagating approximately uniformly. The fact that they \emph{do}
change slightly, with brighter images to the right, indicates that
the feature is gaining intrinsic brightnsss by accumulating material
as it propagates.

The horizontal positions and error bars in Figures \ref{fig:liftoff}
and \ref{fig:Propagation} are the results of manual feature location
of the cusp, with a point-and-click interface. The white error bars
are based on the sharpness of the feature. In the excess brightness
plot, the feature is easy to see but blurs near the top of Figure
\ref{fig:liftoff} due to the higher levels of both photon noise and
motion blur as the feature accelerates to the top of the coronagraph
field of view. The running difference plot highlights fine scale feature
and helps identify the cusp location near the top of the COR-2 field
of view.

\subsection{\label{sub:Acceleration-profile}Acceleration profile}

\begin{figure}[tbh]

\begin{centering}
\includegraphics[width=6in]{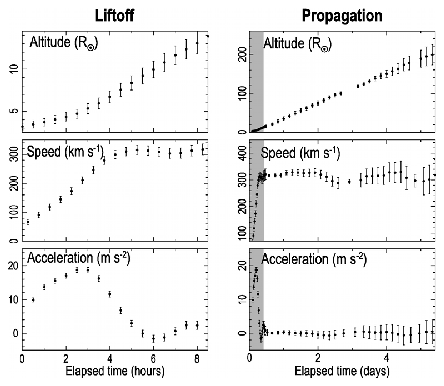}
\par\end{centering}

\caption{\label{fig:accel-plots}Inferred position, speed, and acceleration
of the disconnected plasma from the 2008-Dec-18 event, during onset
(LEFT) and over the full observation period (RIGHT). Error bars are
derived by propagating \emph{a priori} location error and geometric
error in the longitude of the event. The shaded region indicates the
full time range of the left-side plots.}

\end{figure}

Converting angular observed coordinates to examine the inertial behavior
of the plasma requires triangulation using the Law of Sines. Using
the {}``fixed $\Phi"$ approximation (assuming the feature's cusp
is small and that it propagates in a radial line from the Sun), the
feature's radius from the Sun is easily calculated: \begin{equation}
r_{ev}=r_{A}\frac{{sin\left(\epsilon'\right)}}{sin\left(\epsilon\right)},\label{eq:law-of-sines}\end{equation}
where the variables take the meanings in Figure \ref{fig:Observing-geometry}:
$\epsilon$ is the solar elongation of the feature as seen from \emph{STEREO-A},
$\epsilon'$ is the solar elongation of \emph{STEREO-A} as seen from
the feature), and the \emph{STEREO-A} solar distance $r_{A}$ is found
by spacecraft tracking and is supplied by the mission. Although no
camera was present at the event itself, $\epsilon'$ is calculated
by noting that $\epsilon'=180^{\circ}-\epsilon-\left(L-L_{ev}\right)$,
as the feature, \emph{STEREO-A, }and the Sun form a triangle. Figure
\ref{fig:accel-plots} shows the results of the tracking from Figures
\ref{fig:liftoff} and \ref{fig:Propagation}, propagated through
Equation \ref{eq:law-of-sines}. 

As expected, the event rapidly accelerates during the early phase,
reaching a peak acceleration of $20\, m\, s^{-2}$ as the aspect ratio
changes in the initial hours. The acceleration peaks 4-5 hours after
the initial pinch in the streamer belt, or 2-3 hours after the first
observation of a well-formed cusp. The feature reaches its final speed
of $\sim320\pm15\, km\, s^{-1}$ within just 8 hours of the initial
pinch at 04:00 and within 6 hours of the first observation of the
well formed cusp at 06:00, and undergoes no further significant acceleration
nor deceleration during its obsered passage to beyond 1 AU over the
next five days.

\subsection{\label{sub:Mass-profile}Mass profile}

We extracted photometric densities using the feature brightness in
suitable frames. The feature brightness is determined from the density
via the Thomson scattering equation (see, e.g., Howard \& Tappin 2009a
for a clear exposition). Compact features can be treated as nearly
point sources, and the line-of-sight integral for the optically thin
medium reduces to:\begin{equation}
B=B_{\odot}\Omega_{\odot}(r)\sigma_{e}\left(1+cos^{2}\chi\right)\rho\mu_{av}^{-1}d\label{eq:brightness-equation}\end{equation}
where $B$ is the measured feature brightness (in units of emissivity:
$Wm^{-2}SR^{-1}$), $B_{\odot}$ is (still) the solar surface brightness;
$\Omega_{\odot}(r)$ is the solid angle subtended by the Sun at the
point of scatter, well approximated by $\pi r_{\odot}^{2}r_{ev}^{-2}$
everywhere above about 4 $r_{\odot}$; $\sigma_{e}$ is the differential
Thomson scattering cross section, given by half of the square of the
classical electron radius $r_{e}^{2}/2=4.0\times10^{-30}m^{2}$; $\chi$
is the scattering angle (equal to $\epsilon'$ in Figure \ref{fig:Observing-geometry});
$\rho$ is the mass density; $\mu_{av}$ is the average mass per electron
in the coronal plasma; and $d$ is the depth of the feature.

$mu_{av}$ may be calculated from the spectroscopically measured 5\%
He/H number ratio in the corona (Laming \& Feldman 2000) and the
assumption that the helium is fully ionized (yielding two electrons
per ion). This yields $mu_{av}=1.1 m_{p}=1.84\times10^{-27}kg$.

Solving for the line-of-sight integrated mass surface density $\rho d$
gives:\begin{equation}
\rho d=\mu_{av}\frac{B}{B_{\odot}}\Omega_{\odot}^{-1}(r)\sigma_{e}^{-1}\left(1+cos^{2}\epsilon'\right)^{-1}\label{eq:surface-density}\end{equation}
and therefore\begin{equation}
m_{ev}=\left(\rho d\right)wh=\mu_{av}\frac{B}{B_{\odot}}\Omega_{\odot}^{-1}(r)\sigma_{e}^{-1}\left(1+cos^{2}\epsilon'\right)^{-1}\Omega_{ev}^{\left(1+cos^{2}\epsilon'\right)}S^{2}\label{eq:mass}\end{equation}
where $w$ and $h$ are the dimensions shown in Figure \ref{fig:Cartoon}; 
$\Omega_{ev}$ is the solid angle subtended by the feature in
the images; and $S$ is the calculated spacecraft-feature distance,
calculated by the law of sines as for $r_{ev}.$

\begin{figure}[tbh]
\includegraphics[width=5.0in]{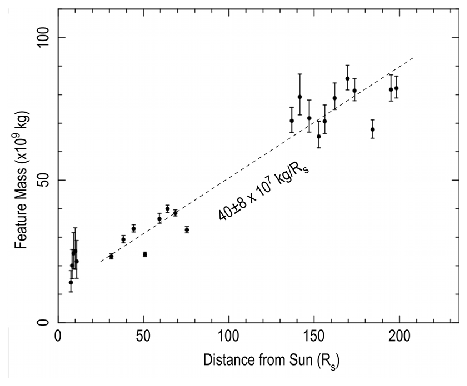}

\caption{\label{fig:mass}Photometrically determined excess mass profile of
the retracting disconnected feature of 2008 Dec 18. Error bars are
based on identification of the feature boundary in the images. The
trendline is extracted from regression of the HI-1 and HI-2 data.
The mass shown is excess mass in the feature compared to the background
solar wind (see text).}

\end{figure}

To extract the mass profile from the data, we generated an image sequence
containing the feature, and marked the locus of the feature visually
using a pixel paint program. Using the generated masks, we summed
masked pixels in the feature for each photometric image, thereby integrating
the feature brightness over the solid angle represented by the corresponding
pixels, to obtain an intensity and an average brightness within the
feature. To account for errors in visual masking, we assigned error
bars based on one-pixel dilation and one-pixel contraction of the
masked locus. We omitted frames with excessive noise, encroachment
of an image boundary, or a star or cosmic ray in or near the feature.
The results of the calculation are given in Figure \ref{fig:mass},
which shows steady accretion of material through most of the journey
through the heliosphere.

Because our photometric analysis is based on subtraction of a calculated
background derived from the data set itself, we measure only excess
feature brightness (not absolute brightness) from Thomson scattering;
thus our brightness measurements and mass estimates are biased low,
because we cannot measure the absolute density of the background.
The initial derived mass of 20-25 Tg translates to an electron number density
of $2\times10^{7}\, cm^{-3}$ in the lower corona, which is comparable
to the density in bright coronal features -- so the total mass may
be up to a factor of order two higher.

The final {}``feature excess'' mass is $8\pm2\times10^{10}kg$, and the
final subtended solid angle is 0.028 SR, for a presented cross-section
of $4.3\pm0.1\times10^{20}m^{2}$. Taking the depth to be the square
root of the observed cross section yields an estimated volume at 1 AU
of $8.9\pm0.3\times10^{31}m^{3}$, for a total estimated excess
electron density of $5\pm1.4\, cm^{-3}$ at 1 AU, which is in good
agreement with slow solar wind densities ($3-10\, cm^{-3}$ when scaled
to 1 AU) that were observed by \emph{Ulysses} in situ in the same
heliographic latitude range (e.g. \citealt{McComas2000}).  Approximately
2/3 of this excess density appears to have been accumulated enroute from
the surroundign solar wind; this is further described in Section \ref{sub:accretion}.

\subsection{Entrained magnetic flux}

\begin{figure}[tbh]
\includegraphics[width=3.0in]{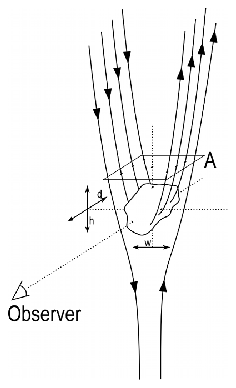}

\caption{\label{fig:Cartoon}Cartoon of the initial acceleration process of
a disconnection event. Tension force along newly released field lines
is balanced by mass entrained on the field lines. By measuring the
acceleration and mass we infer the amount of magnetic flux that was
disconnected.}

\end{figure}

From the mass of the feature, and its acceleration, it is possible to
extract the entrained magnetic field by measuring the rate of change
of momentum and inferring a magnetic tension force via $f=ma$. The
system is sketched in Figure \ref{fig:Cartoon}. The magnetic tension
force is conserved along the open field lines, so we can calculate it
at any convenient cut plane including the one shown. Tension force is
frequently referred to as a ``curvature force'' and calculated
locally; here we integrate around the ``U'', and notice that the
integrated force is just the unbalanced tension on the field lines
contained in the ``U'' shape.  It is therefore given
by
\begin{equation}
f=m_{ev}a_{ev}=f_{B}=\frac{B^{2}A}{2\mu_{0}}=\frac{{\Phi^{2}}}{2\mu_{0}A}\label{eq:tension-force}
\end{equation}
where, here, $B$ is the magnetic field strength (not brightness, as
before). Solving for $\Phi$, 
\begin{equation}
  \Phi=\sqrt{2\mu_{0}dwm_{ev}a_{ev}}\label{eq:flux}
\end{equation}
taking $m_{ev}$ to be $25Tg$ ($2.5\times10^{10}kg$) during the peak of
the acceleration, and taking $w=d=0.2R_{\odot}$ (based on the measured
width of the feature's fork during maximum acceleration, at $6R_{s}$
from the surface ($7R_{s}$ from Sun center) gives
$\Phi=1.6\times10^{11}Wb$ ($1.6\times10^{19}Mx$ ), corresponding to an
average field strength of 8$\mu T$ ( 0.08 Gauss) at that altitude, or
an equivalent $r^{2}$-scaled field of 400$\mu T$ (4 Gauss) at the
surface; this is comparable to accepted values of the open flux
density at the solar surface at solar minimum. Because of the way $m$
was calculated (section 2.5 above) this figure is probably low by a
factor of order $\sqrt{2}$.

\subsection{\label{sub:accretion}Accretion and force balance}

As the disconnectioned structure travels outward, it accretes new
material.  This effect is dramatic: as seen in Figure \ref{fig:mass},
the mass increases by a factor of 3 from the corona to 1 AU.  We
conjecture that the material is accreted by {}``snowplow'' effects
from the plasma ahead of the disconnected cusp as it propagates.  For
the observed mass growth in the feature, new material must be
compressed to become visible in our Thomson scattering images, and the
most plausible way for it to be compressed is via ram effects.  This
scenario also neatly explains the constant speed of the feature, by
balancing the continued tension force from the cusp with accretion
momentum transfer. Here we explore the concept of force balance
between accretion and the tension force, to identify whether some
other model is required in addition to this simple one.

Extending Newton's law to include momentum transfer by accretion,
and neglecting all but the tension force, \begin{equation}
\frac{\Phi^{2}}{2\mu_{0}A_{\Phi}}=m_{ev}a_{ev}+\frac{dm_{ev}}{dt}\Delta v,\label{eq:force}\end{equation}
where the LHS is just the tension force from Equation \ref{eq:tension-force},
with the modification that the cross section of the exiting field
lines is written $A_{\Phi}$; $a_{ev}=0$ after the initial acceleration;
and the second term represents momentum transfer into accreted material,
with$ $ $\Delta v$ being the difference between the feature speed
and surrounding wind speed. The feature is thus in equilibrium between
accretion drag and continued acceleration by the tension force. This
accretion drag is important to the observed increase in feature mass,
because ram pressure against the surrounding wind material is what
compresses incoming material and renders it visible in the data against
the subtracted background.

\begin{figure}[tbh]
\includegraphics[width=5.0in]{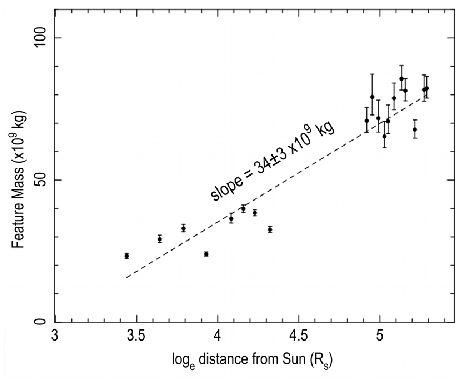}

\caption{\label{fig:Semilog-regression-fit}Semilog regression fit of $m_{ev}(r)$}

\end{figure}

Applying conservation of mass, we can relate $\Delta v$ and the average
density of the background solar wind through which the feature is
propagating:\begin{equation}
\rho_{sw}=\frac{dm_{ev}/dt}{A_{ev}\Delta v}.\label{eq:conservation-of-mass}\end{equation}
where $A_{ev}$ is the geometrical area presented by the feature to
the slow wind ahead of it. Solving Equations \ref{eq:force} and \ref{eq:conservation-of-mass}
to eliminate $\Delta v$ gives\begin{equation}
\rho_{sw}=\left(\frac{dm_{ev}}{dt}\right)^{2}\left(\frac{2\mu_{0}}{\Phi^{2}}\right)\left(\frac{A_{\Phi}}{A_{ev}}\right),\label{eq:background-density}\end{equation}
which gives the background solar wind density in terms of the accumulation
rate of mass in the observed feature, assuming constant outflow for
both the wind and the feature, and acceleration by the tension force.
Given the conservation of mass and the approximately constant speed
of the solar wind, $\rho_{sw}$ falls as approximately $r^{-2}$.
Further, we observe nearly self-similar expansion throughout most
of the heliospheric range, so $A_{\Phi}/A_{ev}$ is constant in that
part of the trajectory -- hence $dm_{ev}/dt$ must also fall as $r^{-1}$
during the approximately constant speed portion of the feature's lifetime.
Using this functional form, we can extract an analytic expresson for
the feature mass versus radius. We introduce the $r^{-1}$ dependence by 
switching from the linear regression used in Figure \ref{fig:mass}, to a 
semi-log regression that assumes $dm_{ev}/d(log_{e}(r)$ to be constant.  
Figure \ref{fig:Semilog-regression-fit} shows such a regression, with the 
result that $dm_{ev}/dr=34\pm3\times10^{9}\left(R_{\odot}/r\right)kgR_{\odot}^{-1}$.
Including the measured outflow speed of $315\pm15\, km\, s^{-1}$,
we find that $dm_{ev}/dt=\left(1AU/r\right)\left(7.1\pm1\times10^{4}kg\, s^{-1}\right)$.

Including all of these values into Equation
\ref{eq:background-density}, together with the average particle mass
from Section \ref{sub:Mass-profile}, yields a background wind numeric
density of $n_{sw}\left(1\, AU\right)$ of $30\pm6\,
cm^{-3}\left(A_{\Phi}/A_{ev}\right)$. From the morphology of the
feature in Figure \ref{fig:Propagation}, we conservatively estimate
$A_{\Phi}/A_{ev}<0.25$, i.e. the forward cross section of the
``horns'' of the vee appears to be well under 1/4 of the cross section
of the vee itself.  This value yields a derived background solar wind
density of $n_{sw}<8\, cm^{-3}$ at 1 AU to maintain the force balance in 
Equation \ref{eq:force}.  That figure is again in line with
the wind measurements from \emph{Ulysses} at $15^{\circ}$ heliographic
latitude (McComas et al. 2000), adding to the plausibility of the
accretion force balance picture. The corresponding mass density limit is
$\rho_{sw}<8\times10^{-19}\, kg\, m^{-3}$ at 1 AU

As a sanity check, we can use this $\rho_{sw}$ limit and Equation
\ref{eq:conservation-of-mass} to find that $\Delta v$ must then
be a few tens of $km\, s^{-1},$ i.e. the background wind speed must
be close to $300\, km\, s^{-1}$. 

We conclude that the picture of force balance between snowplow accretion
and the tension force is at least broadly consistent with the observed
feature, though further study of more events (preferably with corresponding
\emph{in situ} measurements of the feature itself) is necessary.

\[
\]

\section{Discussion}

Using data from STEREO/SECCHI, we have identified and measured the
characteristics of a single flux disconnection event and associated
cusp feature, similar to that discovered by McComas et al. (1992),
from initial detection in the lower corona to distances beyond 1 AU.
The cusp feature is formed in the classic X-point geometry and rapidly
accelerates under the tension force to approximately $320\, km\, s^{-1}$,
which it reaches in under 4 hours at an altitude of approximately
10$R_{\odot}$. Thereafter the feature continues to accumulate mass
but maintains approximately constant speed until it is lost to sight
1.2 AU from the Sun. 

Based on photometry, we are able to estimate the onset mass of the
event as $25\, Tg$ and the entrained flux as $160\, GWb,$ corresponding
to a coronal field strength of $0.08\, G$ and an $r^{2}$-normalized
surface open field of $4\, G$ over the projected surface footprint
of the feature. These estimates are likely low by a factor of order
$\sqrt{2}$, because they make use of feature excess brightness rather
than absolute Thomson-scattered brightness in the coronagraph images;
using polarized-brightness imagery could improve the measurement by
separating the non-transient component of the Thomson scattering signal
from the unwanted F coronal background. 

Because our measurements are all based on morphology and photometry,
we have performed several consistency checks to build confidence in
the calculated parameters of the feature as it propagates. In particular,
a model of simple force balance between the tension force and mass
accretion is consistent with both the inferred magnetic field and
accepted values for background slow solar wind density and speed. 

Simple accretion models such as we developed here demonstrate clearly
why ejected features such as U-loops or CMEs seem frequently to propagate
at near constant speed: under continuous weak driving, an equilibrium
forms rapidly between the driving force and momentum transfer by mass
accretion. The equilibrium outflow speed is the sum of a large, fixed
(or at least driver-independent) speed -- that of the surrounding
wind -- with a smaller offset speed that drives mass accretion. Thus
the feature speed is quite insensitive to the driver. In our case,
doubling the tension force would only increase the outflow rate by
$\sim10\%$. 

The event under study is well presented, but is not unusual at all;
such events are easy to identify in heliospheric image sequences,
because of their distinctive {}``U'' and cusp shape; they are readily
traced back to the corona. This technique represents a new, very effective
way of finding these disconnection events, which are small and hard
to identify\emph{ }in the coronagraph sequences alone, but are strongly
and easily visible in the processed heliospheric images.

In an initial reduced data set of 36 days near the deepest part of
the recent extended solar minimum (2008 Dec -- 2009 Jan), we identified
12 such events; all of them were identified by tracking {}``V''
or {}``U'' shapes back from the heliospheric images to the corona.
Assuming the present feature to be typical, and considering that the
single viewpoint affords clear coverage of about 1/4 of the circumference
of the Sun, we estimate the global disconnection feature rate at that
time to be over $1\, event\, d^{-1}$, and the flux disconnection
rate to thus be at least of order $60\, TWb\, y^{-1}$. Expanded to
a 1 AU sphere, this amounts to a rate of change of the open field
of order $0.2\, nT\, y^{-1}$, which is a significant fraction of
the observed cycle-dependent rate of change of the open heliospheric
field (e.g. Schwadron, Connick, \& Smith 2010). These figures are
based on a single calculated flux and an event rate obtained by initial
visual inspection of a single 36-day data set, and hence are merely
rough estimates -- but they indicate that flux disconnections of this
type are important to the global balance of open flux. Further study,
in the form of a systematic survey, is needed to determine whether
they are the primary mechanism of flux disconnection from the Sun.

\acknowledgements{The authors thank the STEREO instrument teams for making their data
available. Our image processing made heavy use of the freeware Perl
Data Language (http://pdl.perl.org). The work was enhanced by enlightening
conversations with J. Burkepile, C. Eyles, and N. Schwadron, to whom
we are indebted. This work was supported by NASA's SHP-GI program,
under grant NNG05GK14G.}

\section{References}

\bibliographystyle{apj}
\bibliography{ms}

\end{document}